\documentclass[11pt]{article}
\usepackage{epsfig,amsmath,amssymb,euscript}
\usepackage[square,comma,sort&compress,numbers]{natbib}
\usepackage{color}
\usepackage{graphicx}
\usepackage{lineno}
\usepackage{epsfig,amsmath,amssymb,euscript,mathtools}

\newcommand{\Keywords}[1]{\par\noindent 
{\small{\em Keywords\/}: #1}}

\begin{document}
\title{Beyond Strictly Proper Scoring Rules: \\
       The Importance of Being Local}


\author{ Hailiang Du$^{1,2}$ \\[2ex]
        $^1$Department of Mathematical Sciences,\\
   Durham University, Durham, DH1 3LE, U.K.\\[1ex]        
   $^2$Centre for the Analysis of Time Series,\\
   London School of Economics, London WC2A 2AE. UK\\[1ex]
   Email: hailiang.du@durham.ac.uk     
   }

\date{\today}

\maketitle

\begin{abstract}

The evaluation of probabilistic forecasts plays a central role both in the interpretation and in the use of forecast systems and their development. Probabilistic scores (scoring rules) provide statistical measures to assess the quality of probabilistic forecasts. Often, many probabilistic forecast systems are available while evaluations of their performance are not standardized, with different scoring rules being used to measure different aspects of forecast performance. Even when the discussion is restricted to strictly proper scoring rules, there remains considerable variability between them; indeed strictly proper scoring rules need not rank competing forecast systems in the same order when none of these systems are perfect. The locality property is explored to further distinguish scoring rules. The nonlocal strictly proper scoring rules considered are shown to have {a property that can produce ``unfortunate" evaluations. {Particularly the fact that Continuous Rank Probability Score prefers the outcome close to the median of the forecast distribution regardless the probability mass assigned to the value at/near the median raises concern to its use.} The only local strictly proper scoring rules, the logarithmic score, has direct interpretations in terms of probabilities and bits of information. {The nonlocal strictly proper scoring rules, on the other hand, lack meaningful direct interpretation for decision support.} The logarithmic score is also shown to be invariant under smooth transformation of the forecast variable, while the nonlocal strictly proper scoring rules considered may, however, change their preferences due to the transformation. It is therefore suggested that the logarithmic score always be included in the evaluation of probabilistic forecasts.}

\end{abstract}
\Keywords{scoring rule; forecast evaluation; skill score.}

\section{Introduction}

Forecast evaluation has a long history of being a crucial topic for model development and decision support. The outputs from a stochastic model can be naturally interpreted in the form of probabilistic forecast. Given a deterministic model, uncertainty in the initial state due to the observational noise; limited computational power; and model discrepancy prevent one from making a perfect deterministic forecast of the future or even identifying the Truth in the past. In order to account for all sorts of uncertainties, the model outputs are often interpreted as probabilistic forecasts with the aim of providing useful information for decision support. Probabilistic forecasts have been widely adopted in various fields including meteorology, social science, pharmacology, economics and finance; and have become common in operational forecasting over the last quarter century.

The evaluation of probabilistic forecasts plays a central role both in the interpretation and in the use of forecast systems and their development. Such evaluation has not yet been standardized, with many different probabilistic scoring rules~\citep{Gneiting07,Jolliffe03,Roulston02,Wilks95} being used. As probabilistic forecasts become more common, the need to select (probabilistic) scoring rule(s) for constructing probabilistic forecasts, calibrating forecast systems, ranking competing forecast systems and quantifying forecast improvement has {led to} the research work presented in this paper. 

The importance of using strictly proper scoring rules has been noted in the literature~\citep{Brocker07}, {as only strictly proper scoring rules encourage the forecaster to be honest, i.e. reporting a forecast probability distribution gives an optimal expected score only when the verification is, in fact, drawn from that probability distribution.} When the discussion is restricted to strictly proper scoring rules, however, there remains considerable variability between scoring rules (there are, in fact, an infinite number of strictly proper scoring rules). And strictly proper scoring rules need not rank competing forecast systems in the same order when none of these systems are perfect. 

The locality property is explored to further distinguish various strictly proper scoring rules. {A property that {reflects} ``unfortunate"\footnote{Fortuna was the goddess of fortune (luck) in Roman religion. ``Unfortunate" in this paper refers to bad advice, which is a disaster in terms of decision support.} evaluations is introduced. Nonlocal strictly proper scoring rules considered are shown to have a mathematical property, named implausible, that could produce ``unfortunate" evaluations. A few striking examples of the potential issues that result from the use of nonlocal scoring rules are presented. The only local strictly proper scoring rule, the logarithmic score (also known as Ignorance), has direct interpretations in terms of probabilities and bits of information. The nonlocal strictly proper scoring rules are found to lack meaningful direct interpretation. The logarithmic score is also shown to be invariant under smooth transformation of the forecast variable, while the nonlocal strictly proper scoring rules considered may, however, change their preferences due to the transformation.}

This paper emphasizes the fact that being strictly proper is {not sufficient in decision support} when measuring the difference between imperfect forecast systems and suggests that the only local scoring rule, Ignorance, should always be included in the evaluation of probabilistic forecasts.

The definition of a scoring rule for probabilistic forecast and the importance of using strictly proper scoring rules are presented in Section 2. A number of strictly proper scoring rules are defined in Section 3. A common example of strictly proper scoring rules ranking forecast systems differently without the presence of True underlying distribution is given in Section 4. The locality property is defined and discussed in Section 5. {Section 6 introduces a mathematical property that reflects ``unfortunate" evaluations and shows nonlocal scoring rules considered have such property. The interpretation of local and nonlocal scoring rules are discussed in Section 7. Section 8 investigates the behavior of proper scoring rules when smooth transformation is applied to the forecast variable. }Section 8 provides discussion and conclusions.

\section{Probabilistic Scoring Rules and Importance of Being Strictly Proper}

While the true value of a forecast is most clearly reflected in its utility to the end user, probabilistic scores are fundamental to the performance analysis of probabilistic forecasts. Ideally they provide a general measure of future forecast quality, independent of any specific end user~\citep{Brocker07}. 
A probabilistic score (scoring rule) is a function $S(p(x),Y)$, where $p(x)$ is a probability density function and $Y$ is the outcome. In this paper, probabilistic forecasts in the form of probability density functions (PDFs) $p(x)$ are considered\footnote{Results and conclusions presented in this paper also apply to probabilistic forecasts in the form of probability mass functions in the context of categorical variables.}. The notation $p(x)$ denotes the entire function, while $p(Y)$ always denotes the value of the function at the particular outcome $Y$. {By convention, a lower score is taken to reflect a better forecast.} Analytically, the {\it expected score}, 
\begin{equation}
  E(S(p(x),Y))=\int S(p(x),Y)Q(Y)dY,
\end{equation}
which takes the expectation of the scoring rule under the True underlying distribution $Q$ from which the outcome $Y$ is drawn, quantifies the quality of a forecast system.  In practice, an archive of forecast-outcome pairs is required to evaluate the quality of a forecast system. It contains a large number $N$ of forecasts $\{p_i(x), i=1,\ldots,N\}$ and corresponding outcomes $\{Y_i, i=1,\ldots, N\}$. The forecast system yields an {\it empirical score}:
\begin{equation}
  S_{emp}=\frac{1}{N}\sum_{i}^{N} S(p_i(x),Y_i).
\end{equation}
Note the size of the forecast archive can play a major role in determining the significance of the result~\citep{Machete16}, regardless of which scoring rule is employed~\citep{Brocker07}.

Several scoring rules are widely used for the evaluation of probabilistic forecasts \citep{Brier50,Epstein69,Gneiting07,Good52,Jolliffe03,mason-2009,Selten98,Wilks95}; different scoring rules might quantify different attributes of the forecast. Note, however, that Good (1952)'s logarithmic score, also known as Ignorance, (defined below in Section 3d.) is the only scoring rule consistent with the use of (log) likelihoods to evaluate assessors or Bayesian inference~\citep{Winkler69, Winkler96}. 

Since any functional form based on {$p(x)$ and $Y$} could be considered as a scoring rule, one may introduce and use a scoring rule that favors to particular forecast system {which might lead to dishonest and misleading evaluations where the scoring rule encourages the forecaster to select a probabilistic forecast distribution that the forecaster knows is not correct (For example the well-known Finley (1884) tornado forecasts~\citep{Murphy96,Stephenson00}). To avoid such dishonest evaluation, }strictly proper~\citep{Roby65,Toda63} scoring rules (defined in the following) are preferred. The term, {\it proper}, was first introduced by~\cite{Winkler68}, while the general idea goes back to~\cite{Brier50} and~\cite{Good52}. A scoring rule, $S(p(x),Y)$, is said to be {\it proper} if inequality (3) holds for any pair of forecast PDFs, and {\it strictly proper} when equality implies $p = q$: 
\begin{equation}
  \int q(z) S(p(x), z) dz \geq  \int q(z) S(q(x), z) dz .
\end{equation}
For a given forecast $p$, a scoring rule evaluated at the outcome is a random variable with values that depend on the outcome $Y$. Note being strictly proper is a property of the functional form of the scoring rule alone, not of the particular distribution $p(x)$ or $q(x)$. {Strictly proper scoring rules give a probabilistic forecast distribution an optimal expected score only when the outcome is, in fact, drawn from that probability distribution~\citep{Brocker07}.} 
In expectation, a strictly proper scoring rule does not judge any other forecast $p$ to {score better than $q$} as a forecast of $q$ itself. {Note that the interpretation of strictly proper does not, however, require one to believe that the True underlying distribution $Q$ exists.} Strictly proper is a property of the scoring rule; it is neither necessary to assume that $Y$ is drawn from any kind of True distribution nor that any kind of data is to hand. 

The question of whether the employed scoring rule is strictly proper or not can be answered independently of any data being considered~\citep{Brocker07}. Although concerns of hedging are often mentioned\citep{Selten98}, strictly proper scoring rules are preferred even when there is no human involvement, as in parameter {selection} \citep{Du12}. 

{While the importance of using strictly proper scoring rules is well recognized \citep{Brocker07,Brown70,fricker-2013}, researchers often face requests to present results under a variety of scoring rules, both proper and nonproper scoring rules. The fact that nonproper scoring rules like Root Mean Squared Error (RMSE) are still widely used in forecast evaluation often leads to confusion and poorly optimized forecast systems. There have been many discussions regarding the evil of RMSE in the literature (see~\cite{Brocker07,McSharry99,Smith15,Wilks95}), therefore RMSE, which is in fact not a strictly proper scoring rule, will not be considered in this paper.}
\section{Strictly Proper Scoring Rules}

A variety 
of strictly proper scoring rules have been introduced since the 1950s. Some of those widely used are listed below:

\subsection{Energy Score Family}

\cite{Gneiting07} introduced the Energy Score family based on \cite{Szekely03} statistical energy perspective. The Energy Score family $S_{ES}$, is defined as follows\footnote{Negative orientation is applied to the original Energy Score defined by \cite{Gneiting07} so that it is consistent with the convention that a lower score reflects a better forecast.}:
\begin{equation}
	S_{ES}(p(x),Y) = E_{p}\|x-Y\|^{\beta}-\frac{1}{2}E_{p}\|x-x^\prime\|^{\beta},
\end{equation}
where $\beta\in(0,2)$ is a real number; $x$ and $x^\prime$ are independent copies of a random vector with distribution $p$; and $\|\cdot\|$ denotes the Euclidean norm. \cite{Szekely03} and \cite{Gneiting07} show that {the Energy Score is strictly proper relative to the class $\mathbb{P}_{\beta}$, where $\mathbb{P}_{\beta}$ denotes the class of the Borel probability measures $p$ such that $E_p\|x\|^\beta$ is finite.} When $\beta=1$, one obtains:
\begin{equation}
  S_{CRPS}(p(x),Y) = E_{p}\|x-Y\|-\frac{1}{2}E_{p}\|x-x^\prime\|.
\end{equation}
This is equivalent to the well-known Continuous Ranked Probability Score\footnote{{The Continuous Ranked Probability Score is generalized from the Ranked Probability Score (RPS)~\citep{Epstein69, Murphy69} which is widely used to evaluate discrete (categorical) probabilistic forecasts.}} (CRPS)~\citep{Matheson76, Unger1985} (and see~\cite{Baringhaus04,Szekely05,Gneiting07} for the proof of equivalence), where it is the integral of the square of the $L^2$ distance between the cumulative distribution function (CDF) of the forecast $p$ and a {step function at} the outcome~\citep{Epstein69},
\begin{equation}
  S_{CRPS}(p(x),Y) = \int \left( \int^{x}_{-\infty}p(z)dz - H(x-Y) \right)^2 dx,
\end{equation}
\noindent where the Heaviside (step) function H is defined as follows:
\begin{equation}
    H(x) =
    \begin{cases}
      0 &\text{if $x < 0$}\\
      1 &\text{if $x \ge 0$}
    \end{cases}
\end{equation}	
The CRPS was to our knowledge first published by~\cite{Brown74}. It can also be considered as a generalization of the Brier Score~\citep{Brier50} (the Brier Score only applies to binary outcomes~\citep{Matheson76,Murphy70}). For a point forecast, the CRPS is equal to the mean absolute error. In the past decade, the CRPS has been widely used by the atmospheric sciences community~\citep{Goddard2013,Scheuerer14,Zhang2014}. 

\subsection{Power Score Family}

Let $\alpha$ be a real number with $\alpha>1$. The Power Score family~\citep{Selten98} $S_{PS}$, is defined as follows:
\begin{equation}
	S_{PS}(p(x),Y) = -\alpha p(Y)^{\alpha-1}+(\alpha-1)\int p^{\alpha}(z) dz,
\end{equation}
The Power Score family is also strictly proper; this can simply be derived from the derivatives of the expected score~\citep{Selten98}. When $\alpha=2$, one obtains the Proper Linear Score (PLS) (also called the Quadratic Score~\citep{Brier50}):
\begin{equation}
	S_{PLS}(p(x),Y) = -2 p(Y)+\int p^{2}(z) dz,
\end{equation}
PLS derives from the (Naive) Linear Score~\citep{Stael70}, $S_{LS}(p(x),Y) = -p(Y)$, which is not a proper scoring rule as the (Naive) Linear Score favors a $p(x)$ featuring a very small spread and which is centered at the point $x^{\star}$ for which $Q(x^{\star})$ is very large~\citep{Brocker07}.

\subsection{Pseudo-spherical Score Family}

\cite{Good71} introduced the Pseudo-spherical Score family $S_{PSS}$ ($\beta$ is a real number with $\beta>1$), defined as follows:
\begin{equation}
	S_{PSS}(p(x),Y) = -\frac{p(Y)^{\beta-1}}{(\int p^{\beta}(z) dz)^{1/\beta}}.
\end{equation}
The Pseudo-spherical Score family is strictly proper; this can be derived using H{\"o}lder and Minkowski inequality. When $\beta=2$, one obtains the traditional Spherical Score (SPS):
\begin{equation}
	S_{SPS}(p(x),Y) = -\frac{p(Y)}{(\int p^{2}(z) dz)^{1/2}}.
\end{equation}

\subsection{Ignorance}

\cite{Good52} introduced the logarithmic score (also known as Ignorance~\citep{Roulston02}) given by\footnote{Note that defining the logarithmic score in terms of $log_2$ is equivalent to the alternative definition in terms of $ln$ up to the factor $1/ln2$ which does not affect rankings of different forecast systems.}:
\begin{equation}
	S(p(x),Y) = -log_2(p(Y)),
\end{equation}
\noindent where $p(Y)$ is the density assigned to the outcome $Y$. Ignorance (IGN) is a strictly proper scoring rule; this can be derived using Kullback-Leibler inequality~\citep{Kullback51}. {The expected (with respect to $p$) IGN is also a famous information measure, Shannon entropy.} In addition, the expected IGN of $p$ relative to a distribution $q$ becomes the classical Kullback-Leibler divergence~\citep{Kullback59}.

 \section{Different Strictly Proper Scoring Rules Rank Forecast Systems Differently}

Obviously the Energy Score family, Power Score family and Pseudo-spherical Score family contain an infinite number of strictly proper scoring rules. \cite{Jose08} have introduced weighted scoring rules by {blending} the Power Score with the Pseudo-spherical Score; the weighted scoring rule is shown to {be} strictly proper too. Furthermore, \cite{Toda63} proved that a linear transformation of a strictly proper scoring rule is also strictly proper. {Given a strictly proper scoring rule, a forecast system providing $Q$ will always be preferred whenever it is included amongst those under consideration. When none of the competing forecast systems are perfect, then }even strictly proper scoring rules may rank two forecast systems differently, making it impossible to provide definitive statements regarding the relative merit of imperfect forecast systems without considering an additional measure of forecast quality. 

Consider the case where outcomes are independent random draws from a standard Gaussian distribution. Two forecast systems are constructed, where forecast system A uses $N(0,\sigma^2)$ and forecast system B uses $N(0,1/\sigma^2)$ where $\sigma>1$. Obviously neither of the forecast systems is perfect; forecast system A represents a wider distribution around $0$ with larger standard deviation while forecast system B represents a narrower distribution with smaller standard deviation. Figure 1 shows the expectation (under the True distribution, $N(0,1)$) of various scoring rules (Ignorance\footnote{Ignorance is downscaled by a factor of 20 in order to have a similar scale with other scoring rules in Figure 1.}, Continuous Rank Probability Score, Proper Linear Score and Spherical Score) of forecast system A relative to forecast system B as a function of $\sigma$.  If the relative score (also known as {\it skill score}) is negative, {it indicates forecast system A outperforms forecast system B}. Both IGN and PLS prefer wider\footnote{This particular example, based on Gaussian distributions, is designed to show that different scoring rules may rank forecast systems differently and not to indicate whether each of the scoring rules considered here prefers wider or narrower forecast distributions in general, which is not true either.} forecast distribution (forecast system A) than narrower forecast distribution (forecast system B). The CRPS, on the contrary, ranks forecast system B higher than A. Interestingly SPS considers both imperfect forecast system as having the same forecast quality with the expected relative score being zero. (Note given any finite sample of forecasts, there is a $50\%$ chance that {the empirical SPS} prefers forecast A (or B) to the other.) A more thorough investigation in contrasting how certain scoring rules would rank competing forecasts of specified departures from the target distribution can be found in \cite{Machete13}.

\begin{figure}[h]
\centering
{
 \epsfig{file=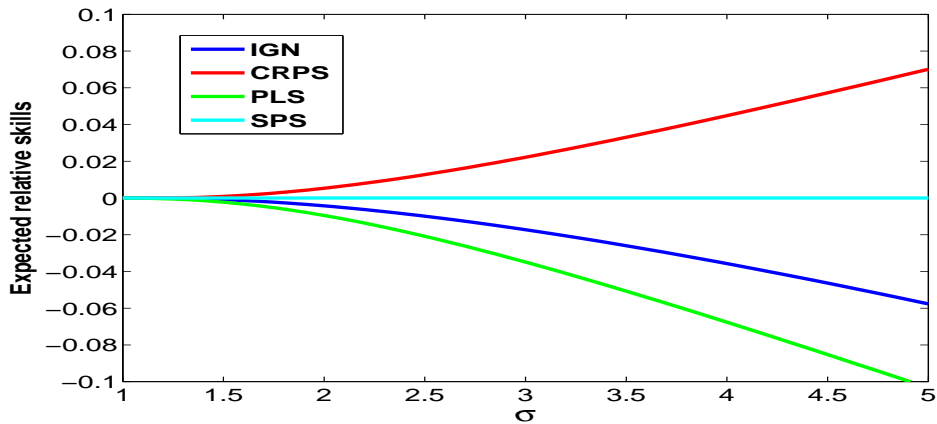, width=0.85\columnwidth, height=6cm}

}
   \caption{The expectation of various scoring rules of forecast system A, $N(0,\sigma^2)$, relative to forecast system B, $N(0,1/\sigma^2)$, where the outcome is drawn from a standard Gaussian distribution. {A negative relative score suggests system A outperforms system B.}}
\label{gmt}
\end{figure}

\section{Locality}

To distinguish between strictly proper scoring rules, the locality property is explored here. A scoring rule is {\it local} if the probabilistic forecast is evaluated only at the actual outcome, which means that the scoring rule depends solely on the probability assigned to the outcome, rather than being rewarded for other features of the forecast distribution, such as its shape.
\cite{Shuford66} and~\cite{Bernardo79} show that every local, smooth and proper scoring rule for continuous variables is equivalent to (an affine function of) IGN, which makes IGN the only proper local scoring rule for continuous variables. Thus all other proper scoring rules, including those listed in Section 3, are nonlocal. The locality property itself does not suggest whether local or nonlocal scoring rules should be preferred, although it might seem unreasonable that features of the forecast other than the value it assigned to the outcome should matter at all. {In the following sections, the preference of a local scoring rule is supported based on both mathematical properties and interpretation of the scoring rule. }

\section{Implausible}

{A mathematical property called implausible is introduced in this section.} Nonlocal scoring rules listed in Section 3 are shown to have such undesirable property; striking ``unfortunate" evaluation examples that result from the use of nonlocal scoring rules are presented below.  

A scoring rule is {\it implausible}\footnote{\cite{Smith15} defined ``perverse" scoring rules to be those which systematically prefer forecasts which place a lower probability on the outcome. \cite{Maynard16} considered a scoring rule to be ``not feasible" when a probable event scores worse than an improbable one. These definitions are elevated here.}, if for ANY $r > 1, r \in \mathbb{R}$, there exist two forecast systems $p_1(x)$ and $p_2(x)$, and $Y$, where $p_1(Y)/p_2(Y)=r$ while $S(p_1,Y)>S(p_2,Y)$. {In other words, an implausible scoring rule means that for all $r > 1$ it is possible to find $p_1(x)$, $p_2(x)$ and $Y$ (which may all vary with $r$) such that $p_1(Y) / p_2(Y) = r$ and $S(p_1, Y) > S(p_2, Y)$. Ignorance is clearly not implausible as given $p_1(Y) / p_2(Y) = r$, $S(p_1, Y)$ would always be smaller than $S(p_2, Y)$ by $\log_2 r$.} The Energy Score family is implausible; this can be shown via investigating an undesirable mathematical property of the Energy Score. Take the derivative of the Energy Score respect to the outcome $Y$ {(where $Y$ is a realization of the random variable $x$)}:
\begin{equation}\label{pl} 
  \frac{\partial S_{ES}(p(x),Y)}{\partial Y}  = \int _{-\infty}^{Y}\beta(Y-x)^{\beta-1}p(x)dx-\int _{Y}^{\infty}\beta(x-Y)^{\beta-1}p(x)dx
\end{equation}
The zero solution of the RHS of Eq. 13 only relies on the location of $Y$ regardless the value of $p(Y)$. For the CRPS, where $\beta=1$, $\min\limits_{Y}S(p,Y)$ is achieved when $\int _{-\infty}^{Y}p(x)dx-\int _{Y}^{\infty}p(x)dx=0$ which gives $Y$ as being the median of $p(x)$. {Such mathematical property} may lead to ``unfortunate" results as illustrated in Figure 2.\footnote{It is usual to analyze how scores change as the forecast varies, while this and the following examples {investigate} how the values of scores change as the outcome varies.} The blue line and red line represent two forecast systems A and B (each based on a Bimodal distribution with the same shape but different centers). Intuitively, one would expect that if the outcome lands between -0.5 and 0.5 {(or more generally that the outcome is drawn from some PDF which is bounded between -0.5 and 0.5) forecast system B shall be preferred as system B would assign significantly more probability mass to the outcome than system A (especially when the outcome lands around 0);} similarly if the outcome lands between 0.5 and 1.5 forecast system A shall be preferred. The green line represents the CRPS of system A relative to system B, a negative (below the dotted zero line) relative score suggests system A outperforms system B according to the CRPS. It appears that if the outcome lands between -0.5 and 0.5, the CRPS would prefer system A over B even when system B assigns significantly more probability mass to the outcome than system A. This is due to the fact that the CRPS prefers the outcome to be close to the median of the forecast distribution no matter how much probability mass is around the median. Obviously the CRPS is implausible, as shown in Figure 2, when $Y=1$, $p_A(Y)/p_B(Y)=\infty$ while $S_{CRPS}(p_A,Y)>S_{CRPS}(p_B,Y)$ (similar examples can be found to show all members of the Energy Score family are implausible). Ironically, if the forecast system A is delivered to the user, the developer of forecast system A would hope the outcome lands at 0 in order to achieve the best CRPS despite the fact the forecast system A assigns 0 probability to the outcome. {Considering a parameter estimation scenario, if the observed outcomes are drawn from a delta function or a sharp Gaussian distribution centered at $0$ and the forecast distribution is a Bimodal distribution with its center to be tuned, tuning the parameter based on the CRPS would converge to a Bimodal distribution centered at $0$ where the probability mass assign to the outcome would always be near $0$.

The example shown in Figure 2. contradicts the claim~\citep{Kohonen06,Boero11,Todter12} that the CRPS/RPS gives credit for assigning high probabilities to the values near but not identical to the outcome. This kind of claims is mostly originated from \citep{Holstein70}, where Sta\"el von Holstein shown that the RPS is ``sensitive to distance" from the ``true" outcome. {Actually the ``sensitive to distance" defined by Sta\"el von Holstein is based on his definition of ``more distant from the true event" (P360 of \citep{Holstein70}), which is, however, NOT equivalent to assigning high probabilities to the values near but not identical to the outcome. It was, in fact, noted by Sta\"el von Holstein himself (section 6 of \citep{Holstein70}) that his definition of ``more distant from the true event" is rather restrictive and changing the definition to an alternative definition \citep{Murphy70b} will lead to the RPS not being ``sensitive to distance", which is consistent with the example shown in Figure 2. }}

\begin{figure}[h]
\centering
{
 \epsfig{file=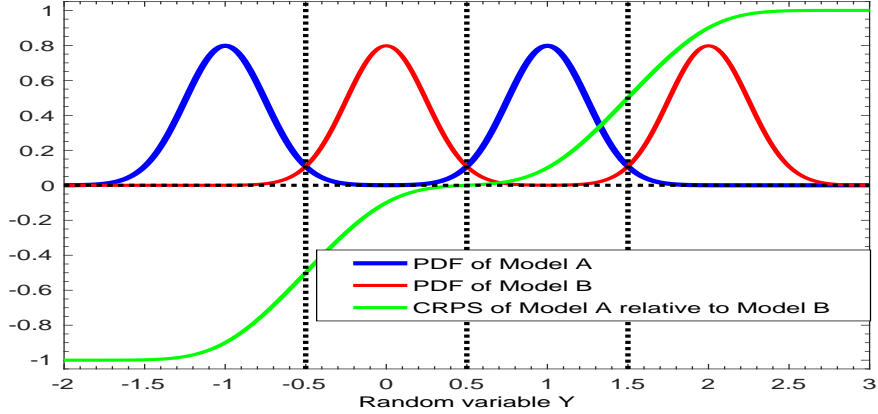, width=0.85\columnwidth, height=6cm}

}
   \caption{Example showing that the Continuous Rank Probability Score produces ``unfortunate" results. The blue line and the red line represent PDFs of forecast systems A and B based on Bimodal distributions with the same shape but different centers. The green line represents the CRPS of system A relative to system B. {A negative relative score suggests system A outperforms system B.} The dashed vertical lines enclose the regions where ``unfortunate" results occur.}
\label{gmt}
\end{figure}

The Power Score and Spherical Score are also implausible. This can be shown in the case, where $p_1(x)$ and $p_2(x)$ are both Gaussian distributions.

Let $p_1(x)$ be a Gaussian distribution with mean $u_1$ and standard deviation $\sigma_1$, then 
\begin{equation}\label{pl}
\begin{multlined}
	 \int^{\infty}_{-\infty} p_1^{\alpha}(z) dz=\int^{\infty}_{-\infty} (\frac{1}{\sqrt{2\pi}\sigma_{1}}e^{-\frac{(z-u_1)^2}{2\sigma_{1}^{2}}})^{\alpha} dz \\
	 = (2\pi)^{\frac{1-\alpha}{2}}\alpha^{-\frac{1}{2}}\sigma_1^{1-\alpha}
\end{multlined}
\end{equation}
Let $p_2(x)$ be a Gaussian distribution with mean $u_2$ and standard deviation $\sigma_2$. To prove the Power Score family is implausible, one needs to find $p_1(\cdot)$, $p_2(\cdot)$ and $Y$ so that $p_1(Y)=rp_2(Y)$ but $S_{PS}(p_1(x),Y)>S_{PS}(p_2(x),Y)$, which requires:
\begin{equation}\label{pl}
\begin{multlined}
	 -\alpha p_1(Y)^{\alpha-1}+(\alpha-1)\int^{\infty}_{-\infty} p_1^{\alpha}(z) dz > -\alpha p_2(Y)^{\alpha-1}+(\alpha-1)\int^{\infty}_{-\infty} p_2^{\alpha}(z) dz \\
	 -\alpha p_1(Y)^{\alpha-1}+(\alpha-1)(2\pi)^{\frac{1-\alpha}{2}}\alpha^{-\frac{1}{2}}\sigma_1^{1-\alpha} > -\alpha p_2(Y)^{\alpha-1}+(\alpha-1)(2\pi)^{\frac{1-\alpha}{2}}\alpha^{-\frac{1}{2}}\sigma_2^{1-\alpha} \\	 
\end{multlined}
\end{equation}
Note that even if $p_2(Y)=0$, it is still possible that $S_{PS}(p_1(x),Y)>S_{PS}(p_2(x),Y)$, as long as $S_{PS}(p_1(x),Y)>0$, as one can always find $\sigma_2$ large enough so that $(\alpha-1)\int^{\infty}_{-\infty} p_2^{\alpha}(z) dz$ is smaller than $S_{PS}(p_1(x),Y)$. To have $S_{PS}(p_1(x),Y) > 0$: 
\begin{equation}\label{pl}
\begin{multlined}
	 p_1(Y)<(\alpha-1)^{\frac{1}{\alpha-1}}\alpha^{-\frac{3}{2(\alpha-1)}}\frac{1}{\sqrt{2\pi}\sigma_1} \\	 
	 p_1(Y)<(\alpha-1)^{\frac{1}{\alpha-1}}\alpha^{-\frac{3}{2(\alpha-1)}}p_1(u_1) \\
\end{multlined}
\end{equation}
This condition also defines a vulnerable subspace where the evaluation using Power Scores might be misinformative. Figure 3 gives an example where PLS may produce ``unfortunate" results. The blue line and red line represent the PDFs of two forecast systems A and B. Intuitively one would expect that if the outcome is less than $-4$ (or between $-2$ and $-1$), system A shall be preferred as system A would assign significantly more probability mass around the outcome than system B. On the contrary, the relative PLS (the green line) prefers system B instead as positive relative PLS would be observed as shown in Figure 3. Ironically, if the forecast systems A and B are delivered to the user, the developer of forecast system B would hope for the outcome being smaller than -4 in order to ``outperform" forecast system A by achieving better PLS despite the fact that forecast system B assigns $\sim0$ probability to the outcome.
\begin{figure}[h]
\centering
{
 \epsfig{file=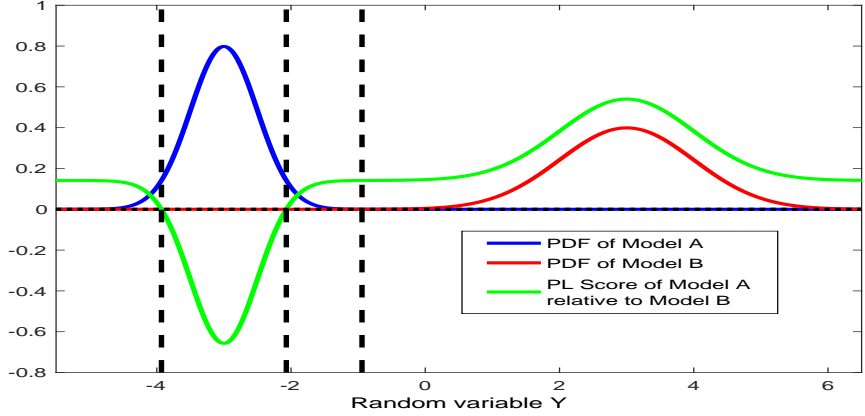, width=0.85\columnwidth, height=6cm}

}
   \caption{Example showing that the Proper Linear Score produces ``unfortunate" result due to the fact it is implausible. The blue line and the red line represent PDFs of forecast system A ($N(-3, 0.5^2)$) and B ($N(3, 1^2)$). The green line represents the PLS of system A relative to system B. {A negative relative score suggests system A outperforms system B.} The left side of the dashed vertical line at $-4$ and the region between dashed vertical lines at $-2$ and $-1$ define the regions where ``unfortunate" results occur.}
\label{gmt}
\end{figure}

Similarly to prove the Pseudo-Spherical Score family is implausible, one needs to find $p_1(\cdot)$, $p_2(\cdot)$ and $Y$ so that $p_1(Y)=rp_2(Y)$ but $S_{PSS}(p_1(x),Y)>S_{PSS}(p_2(x),Y)$, which requires:
\begin{equation}\label{pl}
\begin{multlined}
	 -\frac{p_1(Y)^{\beta-1}}{(\int^{\infty}_{-\infty} p_1^{\beta}(z) dz)^{1/\beta}} > -\frac{p_2(Y)^{\beta-1}}{(\int^{\infty}_{-\infty} p_2^{\beta}(z) dz)^{1/\beta}} \\
	-\frac{p_1(Y)^{\beta-1}}{(2\pi)^{\frac{1-\beta}{2\beta}}\beta^{-\frac{1}{2\beta}}\sigma_1^{\frac{1-\beta}{\beta}}} > -\frac{p_2(Y)^{\beta-1}}{(2\pi)^{\frac{1-\beta}{2\beta}}\beta^{-\frac{1}{2\beta}}\sigma_2^{\frac{1-\beta}{\beta}}} \\
	-\frac{(rp_2(Y))^{\beta-1}}{\sigma_1^{\frac{1-\beta}{\beta}}} > -\frac{p_2(Y)^{\beta-1}}{\sigma_2^{\frac{1-\beta}{\beta}}}  \\	
	\sigma_2  > r^{\beta}\sigma_1 \\
\end{multlined}
\end{equation}
%

Note the condition in Eq. 17 also places a restriction onto $Y$, as $\sigma_2$ gets larger, the maximum value of $p_2(x)$ can be smaller than $\frac{p_1(Y)}{r}$. Therefore $Y$ has to be chosen so that $\frac{p_1(Y)}{r} \leq p_2(u_2)$, i.e. $\frac{p_1(Y)}{r} \leq \frac{1}{\sqrt{2\pi}\sigma_2}$ and as $\sigma_2  > r^{\beta}\sigma_1$, it requires $p_1(Y)< \frac{1}{\sqrt{2\pi}\sigma_1}r^{1-\beta}$. This condition also defines a vulnerable subspace (given $r>1$) where the evaluation using the Pseudo-Spherical Score might be misinformative.

Figure 4 gives an example where SPS may produce ``unfortunate" results. Consider two forecast systems based on Gaussian distributions, where the PDF of system A (blue line) is standard Gaussian and system B (red line) being $N(0, 5^2)$. Intuitively, one would expect that if the outcome lands in the two regions bounded by the black dashed vertical lines, system A shall be preferred as system A would assign significantly more probability mass around the outcome than system B. On the contrary, the relative SPS (the green line) prefers system B instead as positive relative SPS would be observed in both regions. 

\begin{figure}[h]
\centering
{
 \epsfig{file=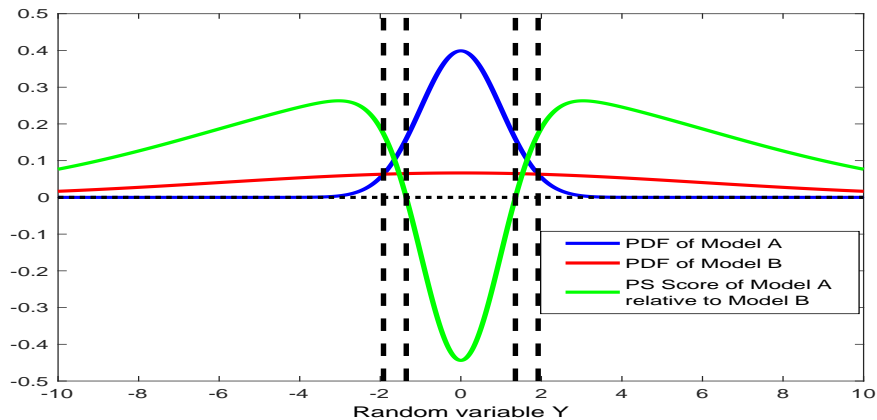, width=0.85\columnwidth, height=6cm}

}
   \caption{Example showing that the Spherical Score produces ``unfortunate" result due to the fact it is implausible. The blue line and the red line represent PDFs of forecast system A (standard Gaussian) and B ($N(0, 5^2)$). The green line represents the SPS of system A relative to system B. {A negative relative score suggests system A outperforms system B.} The dashed vertical lines enclose the regions where ``unfortunate" results occur.}
\label{gmt}
\end{figure}




\section{Score Interpretation}

The difference between two forecast systems is reflected by the difference between their scores. This provides a rank ordering, and thus a preference. {Without any reference, a single score of a forecast system hardly provides any evaluation information, which is why score interpretation should be considered base on the relative score between forecast systems.}
It is also helpful if the relative score has some meaningful interpretation that relates to the benefit of the users. Otherwise it only indicates which forecast system is better, without answering the question of how much better one is {in a meaningful way that adds value to decision support.} For example, Figure 1 does show the IGN, PLS and CRPS' preferences between the two forecast systems; however, the interpretation of the relative score (the y-axis in Figure 1) is also important to decision-makers.   

{A number of meaningful interpretations to proper scoring rules have been identified in the literature. IGN can be interpreted in terms of gambling returns~\citep{Good52,Hagedorn09,Kelly,Roulston02}. Under a Kelly betting scenario\footnote{In a Kelly betting contest \citep{Kelly}, one bets all of one's wealth on every outcome in proportion to the forecast probability of that outcome. More precisely, a fraction $\omega_i$ of ones wealth, where $\omega_i$ is the forecast probability of event $E_i$ occurring, should be wagered on the $i^{th}$ outcome.}, IGN describes the rate at which the forecaster's fortune increases with time. A house setting fair odds~\citep{Frigg14} based on a forecast system with a lower value of a nonlocal scoring rule is expected to lose money to a gambler who places bets based on a different forecast system with a lower IGN. Through its close relation to Shannon's information entropy, IGN is related to the amount of information expected from a forecast~\citep{Roulston02}. IGN can also be easily communicated as an effective interest rate~\citep{Hagedorn09}. Jose and Winkler (2008) show that the Pseudo-spherical score and power score families can be interpreted as profits in certain decision problems. Note all the interpretations listed above are based on some specific scenario in which one can, in fact, define a corresponding utility function to replace the scoring rule. For example in a Kelly betting contest, one can define a utility function that reflects the rate (at which the forecasters' fortune increases with time) and use such utility function to replace IGN for forecast evaluation. {In practice, it is usually not easy to define a relevant utility function based on probabilistic forecasts for the use of decision support.} It is therefore desirable for a scoring rule to have a rather direct and generic interpretation. }

The expected IGN can be written as:
\begin{equation}\label{pl}
\begin{multlined}
  E(S_{IGN}(p(x),Y))=\int \big[ -log_2p(Y) \big]Q(Y) dY
\end{multlined}
\end{equation}
And the expected relative IGN between two probabilistic forecast system $p_1$ and $p_2$ is:
\begin{equation}\label{pl}
\begin{multlined}
   \int \big[ -log_2\frac{p_1(Y)}{p_2(Y)} \big]Q(Y) dY
\end{multlined}
\end{equation}
Therefore the empirical relative IGN score, {$\frac{1}{N}\sum-log_2\frac{p_1(Y)}{p_2(Y)}$}, {reflects} the (average) increase in probability mass that the model forecast $p_1$ placed on the outcome relative to that of the reference forecast $p_2$. Note that although $p_1$ and $p_2$ are probability density functions, $-log_2\frac{p_1(Y)}{p_2(Y)}$ can be interpreted as increase/decrease in probability\footnote{For small values of $\delta$, one can write $P(x<X\leq x+\delta) \approx p(X)\delta$. }, {which gives the Ignorance score a meaningful direct interpretation\footnote{{Note the interpretation of the empirical relative Ignorance score does NOT require the knowledge of the True underlying distribution $Q$.}}}. The relative IGN of two forecast systems also quantifies the information gain (in terms of bits) the model forecast system provides over the reference system. A relative IGN of $1$ bit means that, on average, forecasts from the system assign twice the probability to the outcome compared to the reference forecast~\citep{Roulston02}. 


{Nonlocal scoring rules include contributions from the entire PDF; the scoring rule may be largely determined by outcomes that did NOT occur, making a {meaningful} direct interpretation somewhat challenging.} {For example, the empirical relative PLS between two forecast system $p_1$ and $p_2$ based on a large number N of forecast-outcome pairs is:
\begin{equation}\label{pl}
\begin{multlined}
     [\int p_1^2(z) dz - \int p_2^2(z) dz] + \frac{2}{N}\sum_{i}^{N} [p_2(Y_i)-p_1(Y_i)]
\end{multlined}
\end{equation}
The interpretation of Eq. 20 is clearly more sophisticated than that of the relative IGN. In the second term of Eq. 20, $p_2(Y)-p_1(Y)$, which ranges $(-\infty,\infty$), is the difference between two probability density functions rather than two probabilities. In the context of decision support, it is unclear how to interpret the probability density function(s) meaningfully other than by using $log p_2(Y)-log p_1(Y)$ to reflect the increase/decrease in probability mass placed on $Y$ (this is in fact the approach used by relative IGN). 
The first term of Eq. 20 being a function of the entire PDF of forecast systems (not depending on the outcome $Y$) clearly makes it even more challenge for interpretation. Similar interpretation challenges applies to the CRPS and SPS. There are better ways to interpret these nonlocal scoring rules by using True underlying distribution as a reference. 

}
For example, assuming the True underlying distribution $Q$ exists then the expectation of PLS is:
\begin{equation}\label{pl}
\begin{multlined}
  E(S_{PLS}(p(x),Y))= \int [-2p(Y) + \int p^2(z) dz]Q(Y)dY 
\end{multlined}
\end{equation}
PLS is based on the idea that the scoring rule should reflect ``nearness" of the predicted probability distribution to the True underlying distribution. By straightforward manipulation, it comes to the following representation:
\begin{equation}\label{pl}
\begin{multlined}
  E(S_{PLS}(p(x),Y))=\int [Q(Y)-p(Y)]^{2}dY -\int Q^2(Y) dY  
\end{multlined}
\end{equation}
{The second term in the RHS of Eq. 21 will vanish when comparing two forecast systems using the expected relative score, which gives the expected relative PLS between two probabilistic forecast system $p_1$ and $p_2$:
\begin{equation}\label{pl}
\begin{multlined}
  \int [Q(Y)-p_1(Y)]^{2}dY -\int [Q(Y)-p_2(Y)]^{2}dY  
\end{multlined}
\end{equation}
Therefore the expected relative PLS between two forecast systems can be interpreted with regard to the mean square difference between the forecast distribution and the True underlying distribution $Q$. }

Similarly, the expectation of CRPS can be written as:
\begin{equation}\label{pl}
\begin{multlined}
  E(S_{CRPS}(p(x),Y))=\int [G(Y)-F(Y)]^{2}dY -\int G(Y)(1-G(Y) dY,
\end{multlined}
\end{equation}
where $F(\cdot)$ is the CDF of the forecast distribution and $G(\cdot)$ is the True underlying CDF. The expectation of the relative CRPS between two forecast systems can be interpreted with regard to the mean square difference between the forecast CDF and the CDF of the Truth.

The expected SPS can be written as:
\begin{equation}\label{pl}
\begin{multlined}
  E(S_{SPS}(p(x),Y))=(\int Q(Y)^2 dY)^{1/2} \frac{\int p(Y)Q(Y) dY}{(\int Q(Y)^2 dY)^{1/2}(\int p(Y)^2 dY)^{1/2}}
\end{multlined}
\end{equation}
It can be interpreted regarding the interior angle of deviation between the forecast distribution $p$ and the True underlying distribution $Q$. 

In some cases it makes sense to consider an integration over the True underlying distribution $Q$. The interpretation of the expected relative score with respect to $Q$ is cloudy in reality, for example in weather-like forecasting scenarios, where the same $Q$ distribution is never seen twice over the lifetime of the system. In practice, the True underlying distribution is rarely (if ever) available to provide such an interpretation, and were it to be the use of imperfect probabilistic forecast is mute. {Furthermore, functions of the entire forecast distribution, as in Eq 22. 24. $\&$ 25. can hardly be interpreted in a meaningful way for decision support. Therefore, even if the True underlying distribution were available, it is unclear the interpretations of relative scores derived from Eq 23-25 are informative to the decision maker except providing their preference between two forecast systems.} 

{
\section{Scoring Rules Under Transformation}
In practice, it is common that the variable of interest is not the variable observed but a function of the observed variable. For example, wind power is a function of wind speed cubed; wave power is principally a function of wave height squared~\citep{Savenkov09}. It is desirable for a scoring rule to provide coherent evaluations before and after a smooth transformation being applied to the forecast variable. 
Consider $x^*=\phi(x)$ as a smooth one-to-one (transformation) function of a random variable $x$. The forecast PDF of $x$, $p(x)$, becomes $p(\phi^{-1}(x^*))\frac{d\phi^{-1}(x^*)}{x^*}$ for the random variable $x^*$ after the transformation and the scoring rule $S(p(x),Y)$ becomes $S(p(\phi^{-1}(x^*))\frac{d\phi^{-1}(x^*)}{x^*},Y^*)$, where $Y^*=\phi(Y)$. It is almost certain that the value of a scoring rule will change after the transformation. Note score interpretation should always based on the relative score between forecast systems instead of a single score of a forecast system in order to provide useful information for decision support. It is therefore of interest to investigate whether the relative score will change after taking the transformation and if so, will the scoring rule's preference change as well. Given the relative score between two probabilistic forecast system $p_1$ and $p_2$ by:
\begin{equation}\label{rp}
\begin{multlined}
    S(p_1(x),Y)-S(p_2(x),Y),
  \end{multlined}
\end{equation}
after taking a transformation $\phi(x)$ it becomes
\begin{equation}\label{rpt}
\begin{multlined}
    S(p_1(\phi^{-1}(x^*))\frac{d\phi^{-1}(x^*)}{dx^*},Y^*)-S(p_2(\phi^{-1}(x^*))\frac{d\phi^{-1}(x^*)}{dx^*},Y^*).
  \end{multlined}
\end{equation}
Note that $Y$ and $Y^*$ are one-to-one and $p(x)$ and $p(\phi^{-1}(x^*))\frac{d\phi^{-1}(x^*)}{x^*}$ reflect the same information of a forecast system. Therefore if \eqref{rp} does not equal to \eqref{rpt} based on some scoring rule S, the scoring rule will have a non-unique interpretation of the relative skill between two competing forecast systems. Furthermore if \eqref{rp}$\times$\eqref{rpt} $<0$ for some $Y$ and $\phi$, it indicates such scoring rule might also change its preference due to the transformation, then the use of such scoring rule as an evaluation tool for decision support is questionable. 

Ignorance score is {\it invariant} under smooth transformation as \eqref{rp} and \eqref{rpt} are equal for any smooth transformation, proved in the following:{
\begin{equation}\label{inv}
\begin{multlined}
    S(p_1(\phi^{-1}(x^*))\frac{d\phi^{-1}(x^*)}{x^*},Y^*)-S(p_2(\phi^{-1}(x^*))\frac{d\phi^{-1}(x^*)}{x^*},Y^*) \\
    =-\log p_1(\phi^{-1}(x^*))\frac{d\phi^{-1}(x^*)}{x^*}\bigg\rvert_{Y^*} + \log p_2(\phi^{-1}(x^*))\frac{d\phi^{-1}(x^*)}{x^*}\bigg\rvert_{Y^*}  \\
    =-\log p_1(Y) + \log p_2(Y)=S(p_1(x),Y)-S(p_2(x),Y).
  \end{multlined}
\end{equation}
}
For nonlocal scoring rules like Proper Linear Score, Spheric Score and Continuous Rank Probability score, smooth transformation not only have impact on the value of the relative scores but also may cause the change of their preference. Figure 5 gives an example where the CRPS may contradict itself by changing its preference under transformation (similar examples can be found for PLS and SPS). Following Figure 2, Figure 5(a) compares two forecast systems based on a Bimodal distribution with the same shape but different centers. The green line represents the CRPS of system A relative to system B, a negative relative score suggests system A outperforms system B according to the CRPS. The black dashed vertical line in Figure 5(a) corresponds to the threshold $Y=11.5$, where the CRPS prefers forecast system A when $Y<11.5$ and prefers forecast system B when $Y>11.5$. Figure 5(b) compares the same two forecast systems after cubic transformation being applied to the forecast variable. Clearly the relative CRPS has changed after the cubic transformation. Let $x$ refers to wind speed, then $x^3$ reflects wind power. When the observed wind speed is 10, the relative CRPS (forecast system A relative to forecast system B) is roughly $-0.9$ as in Figure 5(a). Comparing the same\footnote{In fact the information presented by forecast systems A and B remain the same although their pdf changed when the forecast variable wind speed transformed to wind power.} forecast system A and B in terms of wind power under the same observation (wind speed 10 corresponds to wind power 1000), however, the relative CRPS\footnote{The relative CRPS (the green curve) in Figure 5(b) is scaled down by $1.6\times10^5$ in order to have similar magnitude as the PDFs.} as in Figure 5(b) becomes roughly $340$, which indicates the interpretation of CRPS evaluation for comparing forecast system A and B based on a unique observed wind speed is not unique.  Furthermore, the CRPS may even change its preference after the transformation as the threshold (the black solid vertical line in Figure 5(b)) that distinguish the CRPS preference is $Y^3=1700$ rather than $Y^3=11.5^3=1520.875$ (dashed vertical line). If the underlying distribution of the wind speed were bounded between the black dashed line and black solid line, the CRPS would prefer forecast system A before the cubic transformation when the wind speed is evaluated directly, while it prefer forecast system B after the cubic transformation when the wind power (which corresponds to the same wind speed) is evaluated. 


\begin{figure}[!h]

\hbox{
  \epsfig{file=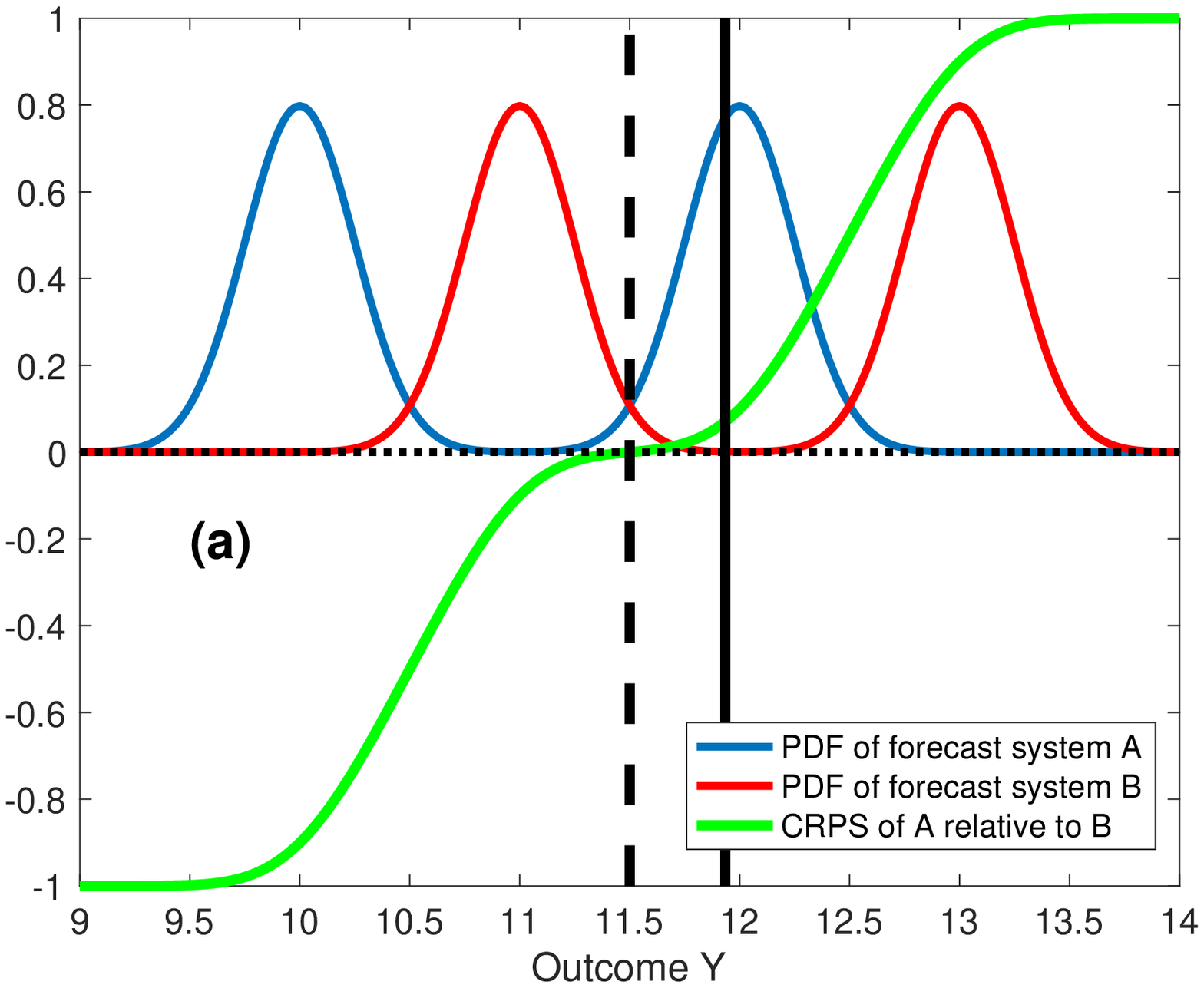, width=0.48\columnwidth, height=6cm}
  \epsfig{file=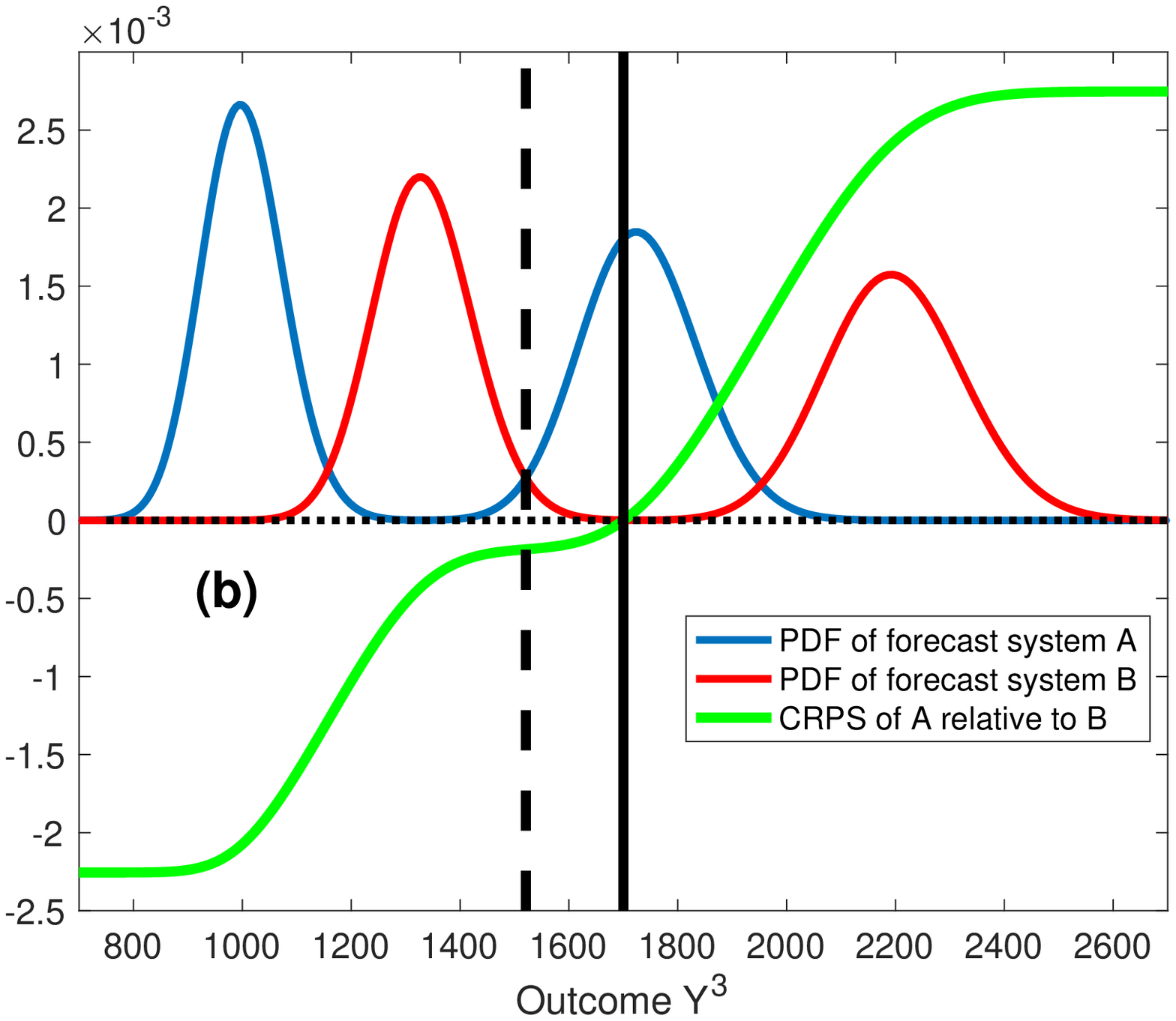, width=0.48\columnwidth, height=6cm}
}  
  
\caption{{Example showing that the Continuous Rank Probability Score changes its preference under transformation. The blue line and the red line represent PDFs of forecast systems A and B. The green line represents the CRPS of system A relative to system B. A negative relative score suggests system A outperforms system B. (a) before the cubic transformation; (b) after the cubic transformation. The black dashed vertical line and solid line in (a), where $Y=11.5$ and $Y=11.935$ respectively, corresponds to those in (b), where $Y=11.5^3$ and $Y=11.935^3$.}}
\label{fig:maperr1}
\end{figure}
}

\section{Discussion and Summary}

Measures of skill play a critical role in the development, deployment and application of probabilistic forecasts. The property of some common strictly proper scoring rules have been discussed. Given a strictly proper scoring rule, the True forecast system will always be preferred whenever it is included amongst those under consideration. In practice, to correctly measure the difference between imperfect forecast schemes, being strictly proper is not enough, as strictly proper scoring rules need not rank competing forecast systems in the same order when none of the forecast systems are perfect. In general, any scoring rules can be presented with the form:
\begin{equation}
	S(p(x),Y) = s_{1}(p(x))+s_{2}(p(x), Y)+s_{3}(p(Y)).
\end{equation}
For local scoring rules, the first two terms in the RHS of Eq. 28 are both zero with only the presence of $s_{3}(p(Y))$, for example the only local proper scoring rule, the logarithmic score (Ignorance). Nonlocal scoring rules contain at least one of the first two terms, for example the Energy Scores consist of $s_{1}$ and $s_{2}$, the Power Scores $s_{1}$ and $s_{3}$ and the Pseudo-sphere Scores only $s_{2}$. The presence of $s_1$ or $s_2$ or both allows the scoring rule to give extra credit to the structure of the forecast PDF. Note such extra credit is not necessarily given for assigning high probabilities to the values near the outcome as the examples in Figures 2-4 show. Without knowing the True underlying distribution, the justification of giving such extra credit is untenable. Nonlocal strictly proper scoring rules considered\footnote{{The author does believe that all non-local scoring rules are implausible, but as there is no general mathematical function of non-local scoring rules, only the popular nonlocal scoring rules (and their families) are considered and shown to be implausible in this manuscript.}} are shown to have property that can produce ``unfortunate" evaluations due to the fact that contributions from the entire shape of the PDF may overwhelm that from the probability assigned to the outcome.  {Particularly the fact that Continuous Rank Probability Score prefers the outcome close to the median of the forecast distribution regardless the probability mass assigned to the value at/near the median raises concern to the use of Continuous Rank Probability Score.} {Ignorance has direct interpretations in terms of probabilities and bits of information while the {}direct interpretation of nonlocal strictly proper scoring rules on the other hand relies on information regarding the unknown (if it even exists) True underlying distribution {as a reference}. The nonlocal strictly proper scoring rules considered may also contradict themselves when a smooth transformation is applied to the forecast variable while IGN is shown to be invariant under smooth transformation.} It is suggested that Ignorance should always be included in the evaluation of probabilistic forecasts.

One of the reasons for using nonlocal scoring rules is to address particular problems where a local scoring rule is not considered ``suitable". For example, Ignorance is infinity if the forecast assigns vanishing probability to an event that obtains. \cite{Selten98} emphasizes that the use of Ignorance implies the value judgment that small differences between small probabilities should be taken very seriously and that wrongly describing something extremely improbable as having zero probability is ``an unforgivable sin". \cite{Roulston02} pointed out that forecasters should replace zero forecast probabilities with small probabilities based on the uncertainties in the forecast PDF. Not to do so means reporting the improbable as the impossible. Within the Bayesian framework, Cromwell's rule states that the use of prior probabilities of 0 or 1 should be avoided. Assigning zero probability to events that are possible {also contradicts} to Laplace's rule of succession~\citep{Jaynes03}. In the insurance sector, the premium is inversely proportional to the probability of an event occurring; zero probability would suggest free insurance. 

In this manuscript, the value outcome is assumed to be certain. In the presence of uncertainty in the value of the outcome (for example due to measurement error), one may obtain benefit by assigning probability to the events that do not match the outcome exactly. Note again this does not imply one should use nonlocal scoring rules, as for nonlocal scoring rules the contributions from the entire shape of the PDF are not designed to account for the uncertainty in the value of the outcome. 
For a local scoring rule, the evaluation can still be considered over the observational uncertainty distribution of the outcome, for example by coupling the forecast distribution with the distribution of observational noise~\citep{Brocker07}.

Scoring rules are designed to assess (probabilistic) forecast performance, which hopefully leads to better decision making. \cite{Bernardo00} argue that a local proper score should be preferred for `pure inference' problems in which the outcome is the sole arbiter of forecast quality, yet there are other forms of scoring rules that would typically be appropriate in more directly practical contexts (see stock control example in \cite{Bernardo00}). Note that in such `more directly practical contexts', if a utility function based on probabilistic forecasts can be conveniently defined according to the practical objective (which often is not the case in practice), there is no need for any kind of scoring rules (using the utility function directly will serve the purpose of forecast evaluation sufficiently). {Any scoring rules can be directly considered as a utility function, yet the meaning of the corresponding utility function relies on the direct interpretation of the skill of the scoring rule. It is questionable whether nonlocal scoring rules can provide any meaningful direct interpretation.} \cite{Smith19} claims that interpretation is a critical aspect in accepting a scoring rule for use in practice; he uses valued property of probabilistic forecasts to support this assertion.

\section*{Acknowledgment}

This research was supported the EPSRC-funded Uncertainty analysis of hierarchical energy systems models: Models versus real energy systems (EP/K03832X/1) and Centre for Energy Systems Integration (EP/P001173/1). Additional support was also provided by Evaluating Probability Scores for the Insurance Sector funded by LSE KEI and Lighthill Risk Network. The author would like to thank Leonard A. Smith and Edward Wheatcroft for reading earlier versions of this article and giving useful feedback; and two anonymous reviewers whose comments and suggestions helped improve and clarify this article.


\begin{thebibliography}{990}

\bibitem{Baringhaus04}
Baringhaus, L., and C. Franz, 2004: On a new multivariate two-sample test. Journal of Multivariate Analysis, 88 (1), 190-206.

\bibitem{Bernardo79}
J. M. Bernardo. Expected information as expected utility. Ann. Stat., 7:686-690, 1979.

\bibitem{Bernardo00}
Bernardo, J. M., and A. F. M. Smith, Bayesian Theory. Wiley, 2000.

\bibitem{Boero11}
Boero, G., J. Smith, and K. F. Wallis, 2011: Scoring rules and survey density forecasts. International Journal of Forecasting, 27, 379-393.

\bibitem{Brier50}
Brier, G.W., 1950: Verification of forecasts expressed in terms of probability. Mon. Wea. Rev., 78, 502 1-3.

\bibitem{Brocker07}
J. Brocker and L.A. Smith. Scoring probabilistic forecasts: On the importance of being proper. Wea. Forecasting, 22:382-388, 2007.

\bibitem{Brown70}
Brown, T. A., 1970: Probabilistic forecasts and reproducing scoring systems. Technical Report RM-6299-ARPA, RAND Corporation.

\bibitem{Brown74}
Brown, T. A., 1974: Admissible scoring systems for continuous distributions. Manuscript P-5235, The Rand Corporation.

\bibitem{Du12}
Du, H., and L. A. Smith, 2012: Parameter estimation using ignorance. Physical Review E, 86, 016213.

\bibitem{Shuford66}
E. H. Shuford, H. E. M., A. Albert, 1966: Admissible probability measurement procedures. Psychometrika, 31.

\bibitem{Epstein69}
Epstein, E. S., 1969: A scoring system for probability forecasts of ranked categories. Journal of Applied Meteorology, 8, 985-987.

\bibitem{fricker-2013}
Fricker, T. E., C. A. T. Ferro, and D. B. Stephenson, 2013: Three recommendations for evaluating climate prediction. Meteorological Applications, 20, 246-255.

\bibitem{Frigg14}
Frigg, R., S. Bradley, H. Du, and L. Smith, 2014: Laplace's demon and the adventures of his apprentices. Philosophy of Science, 81, 31-59.

\bibitem{Gneiting07}
Gneiting, T., and A. E. Raftery, 2007: Strictly proper scoring rules, prediction and estimation. Journal of the American Statistical Association, 102, 477:359-378.

\bibitem{Goddard2013}
Goddard, L., and Coauthors, 2013: A verification framework for interannual-to-decadal predictions experiments. Climate Dynamics, 40, 1-2: 245-272.

\bibitem{Good52}
I. J. Good. Rational decisions. Journal of the Royal Statistical Society, XIV(1), 1952.

\bibitem{Good71}
Good, I. J., 1971: Comment on 'Measuring information and uncertainty' by Robert J. Buehler, 337-339. Holt, Rinehart and Winston, Toronto.

\bibitem{Hagedorn09}
R. Hagedorn and L. A. Smith. Communicating the value of probabilistic forecasts with weather roulette. Meteor. Appl., 16:143155, 2009.

\bibitem{Jaynes03}
Jaynes, E. T., 2003: Probability Theory: The Logic of Science. Cambridge University Press.

\bibitem{Jolliffe03}
Jolliffe, I. T., and D. B. Stephenson, 2003: Forecast Verification: A Practitioner's Guide in Atmospheric Science. Wiley.

\bibitem{Kelly}
Kelly, J. L., 1956: A new interpretation of information rate. Bell System Technical Journal, 35, 917-926.

\bibitem{Kohonen06}
Kohonen, J., and J. Suomela, 2006: Lessons learned in the challenge: Making predictions and scoring them. Machine Learning Challenges. Evaluating Predictive Uncertainty, Visual Object Classification, and Recognising Tectual Entailment, J. Quinonero-Candela, I. Dagan, B. Magnini, and F. d'AlcheBuc, Eds., Springer Berlin Heidelberg, Berlin, Heidelberg, 95-116.

\bibitem{Kullback59}
Kullback, S., 1959: Information Theory and Statistics. Wiley.

\bibitem{Kullback51}
Kullback, S., and R. A. Leibler, 1951: On information and sufficiency. Ann.Math. Stat., 22, 79-86.

\bibitem{Machete13}
Machete, R., 2013: Contrasting probabilistic scoring rules. Journal of Statistical Planning and Inference, 143.

\bibitem{Machete16}
Machete, R., and L. Smith, 2016: Demonstrating the value of larger ensembles in forecasting physical systems. Tellus A, 68, 28 393.

\bibitem{mason-2009}
Mason, S. J., and A. P. Weigel, 2009: A generic forecast verification framework for administrative purposes. Monthly Weather Review, 137, 331-349.

\bibitem{Matheson76}
Matheson, J. E., and R. L. Winkler, 1976: Scoring rules for continuous probability distributions.  Management Science, 22, 1087-1096.

\bibitem{Maynard16}
Maynard, T., 2016: Extreme insurance and the dynamics of risk. Ph.D. thesis, The London School of Economics and Political Science, London, UK.

\bibitem{McSharry99}
McSharry, P., and L. A. Smith, 1999: Better nonlinear models from noisy data: attractors with maximum likelihood. Physical Review Letters, 83.

\bibitem{Murphy69}
Murphy, A. H., 1969: On the 'Ranked Probability Score'. Journal of Applied Meteorology, 8 (6), 988-989.

\bibitem{Murphy70b}
Murphy, A. H., 1970: The Ranked Probability Score and The Probability Score: A Comparison. Monthly Weather Review, 98 (12), 917-924.

\bibitem{Murphy96}
Murphy, A. H., 1996: The finley affair: A signal event in the history of forecast verification. Weather and Forecasting, 11 (1), 3-20.

\bibitem{Murphy70}
Murphy, A. H., and R. L. Winkler, 1970: Scoring rules in probability assessment and evaluation. Acta. Psychol., 34, 273-286.

\bibitem{Roby65}
Roby, T. B., 1965: Belief states: A preliminary empirical study. Behavioral Science, 10.

\bibitem{Roulston02}
M. S. Roulston and L. A. Smith. Evaluating probabilistic forecasts using information theory. Mon. Wea. Rev., 130:1653-1660, 2002.

\bibitem{Savenkov09}
Savenkov, M., 2009: On the truncated weibull distribution and its usefulness in evaluating the theoretical capacity factor of potential wind (or wave) energy sites. University Journal of Engineering and Technology, 1, 21-25.

\bibitem{Scheuerer14}
Scheuerer, M., 2014: Probabilistic quantitative precipitation forecasting using ensemble model output statistics. Quarterly Journal of the Royal Meteorological Society, 140, 1086-1096.

\bibitem{Selten98}
Selten, R., 1998: Axiomatic characterization of the quadratic scoring rule. Experimental Economics, 1.

\bibitem{Smith19}
Smith, L., 2020: Necessary conditions for scoring probability forecast system. in preparation.

\bibitem{Smith15}
Smith, L. A., E. B. Suckling, E. L. Thompson, T. Maynard, and H. Du, 2015: Towards improving the framework for probabilistic forecast evaluation. Climatic Change, 132 (1), 31-45.

\bibitem{Holstein70}
Stael von Holstein, C.-A. S., 1970a: A family of strictly proper scoring rules which are sensitive to distance. Journal of Applied Meteorology, 9 (3), 360-364.

\bibitem{Stael70}
Stael von Holstein, C.-A. S., 1970b: Measurement of subjective probability. Acta Psychologica, 34, 146-159.

\bibitem{Stephenson00}
Stephenson, D. B., 2000: Use of the 'odds ratio' for diagnosing forecast skill. Weather and Forecasting, 15 (2), 221-232.

\bibitem{Szekely03}
Szekely, G. J., 2003: E-statistics: The energy of statistical samples. Technical Report 2003-16,  Bowling Green State University.

\bibitem{Szekely05}
Szekely, G. J., and M. L. Rizzo, 2005: A new test for multivariate normality. Journal of Multivariate Analysis, 93 (1), 58-80.

\bibitem{Toda63}
Toda, M., 1963: Measurement of subjective probability distributions. esd-tdr-63-407. Decision Sciences Laboratory, Electronic Systems Division, Air Force Systems Command.

\bibitem{Todter12}
Todter, J., and B. Ahrens, 2012: Generalization of the Ignorance Score: Continuous Ranked Version and Its Decomposition. Monthly Weather Review, 140 (6), 2005-2017.

\bibitem{Unger1985}
Unger, D. A., 1995: A method to estimate the continuous ranked probability score. Proceedings of the ninth conference on probability and statistics, American Meteorological Society, Boston, USA, 206-213.

\bibitem{Jose08}
V. R. R. Jose, R. F. N., and R. L. Winkler, 2008: Scoring rules, generalized entropy, and utility maximization. Operations Research, 56.

\bibitem{Wilks95}
Wilks, D. S., 1995: Statistical Methods in the Atmospheric Sciences: An Introduction. International Geophysics Series. Academic Press.

\bibitem{Winkler69}
Winkler, R. L., 1969: Scoring rules and the evaluation of probability assessors. J. Amer. Statist. Assoc., 64.

\bibitem{Winkler96}
Winkler, R. L., 1996: Scoring rules and the evaluation of probabilities. Test, 5.

\bibitem{Winkler68}
Winkler, R. L., and A. H. Murphy, 1968: 'good' probability assessors. Journal of Applied Meteorology, 7 (5), 751-758.

\bibitem{Zhang2014}
Zhang, Y., J. Wang, and X. Wang, 2014: Review on probabilistic forecasting of wind power generation. Renewable and Sustainable Energy Reviews, 32, 255-270.




















































\end{thebibliography}
\end{document}